\documentclass[12pt]{article}
 \usepackage{epsfig}
 \def\be{\begin{equation}}
 \def\ee{\end{equation}}
 \def\bea{\begin{eqnarray}}
 \def\eea{\end{eqnarray}}
 \usepackage{graphicx}

 \catcode`\@=11
 \def\lsim{\mathrel{\mathpalette\@versim<}}
 \def\gsim{\mathrel{\mathpalette\@versim>}}
 \def\@versim#1#2{\vcenter{\offinterlineskip
 \ialign{$\m@th#1\hfil##\hfil$\crcr#2\crcr\sim\crcr } }}
 \catcode`\@=12

 \catcode`@=12
\begin{document}
 \thispagestyle{empty}
 \begin{flushright}
 UCRHEP-T578\\
 June 2017\
 \end{flushright}
 \vspace{0.6in}
 \begin{center}
 {\LARGE \bf Dark Gauge U(1) Symmetry for\\
 an Alternative Left-Right Model\\}
 \vspace{1.0in}
 {\bf Corey Kownacki, Ernest Ma, Nicholas Pollard, Oleg Popov, and 
Mohammadreza Zakeri\\}
 \vspace{0.2in}
 {\sl Department of Physics and Astronomy,\\ 
 University of California, Riverside, California 92521, USA\\}
 \end{center}
 \vspace{1.0in}

\begin{abstract}\
An alternative left-right model of quarks and leptons, where the $SU(2)_R$ 
lepton doublet $(\nu,l)_R$ is replaced with $(n,l)_R$ so that $n_R$ is 
not the Dirac mass partner of $\nu_L$, has been known since 1987.  
Previous versions assumed a global $U(1)_S$ symmetry to allow $n$ to be 
identified as a dark-matter fermion.  We propose here a gauge
extension by the addition of extra fermions to render the model free of 
gauge anomalies, and just one singlet scalar to break $U(1)_S$.  This 
results in two layers of dark matter, one hidden behind the other.
\end{abstract}

\newpage
 \baselineskip 24pt

\noindent \underline{\it Introduction}~:\\
The alternative left-right model~\cite{m87} of 1987 was inspired by the $E_6$ 
decomposition to the standard $SU(3)_C \times SU(2)_L \times U(1)_Y$ 
gauge symmetry through an $SU(2)_R$ which does not have the conventional 
assignments of quarks and leptons.  Instead of $(u,d)_R$ and $(\nu,l)_R$ 
as doublets under $SU(2)_R$, a new quark $h$ and a new lepton $n$ per 
family are added so that $(u,h)_R$ and $(n,e)_R$ are the $SU(2)_R$ 
doublets, and $h_L$, $d_R$, $n_L$, $\nu_R$ are singlets.

This structure allows for the absence of tree-level flavor-changing 
neutral currents (unavoidable in the conventional model), as well as 
the existence of dark matter.  The key new ingredient is a $U(1)_S$ 
symmetry, which breaks together with $SU(2)_R$, such that a residual 
global $S'$ symmetry remains for the stabilization of dark matter. 
Previously~\cite{klm09,klm10,bmw14}, this $U(1)_S$ was assumed to be global.  
We show in this 
paper how it may be promoted to a gauge symmetry.  To accomplish this, 
new fermions are added to render the model free of gauge anomalies. 
The resulting theory has an automatic discrete $Z_2$ symmetry which is 
unbroken, as well as the global $S'$, which is now broken to $Z_3$. 
Hence dark matter has two components~\cite{cmwy07}.  They are identified 
as one 
Dirac fermion (nontrivial under both $Z_2$ and $Z_3$) and one complex 
scalar (nontrivial under $Z_3$).

\noindent \underline{\it Model}~:\\
The particle content of our model is given in Table 1, where the scalar 
$SU(2)_L \times SU(2)_R$ bidoublet is given by 
\begin{equation}
\eta = \pmatrix{\eta^0_1 & \eta_2^+ \cr \eta_1^- & \eta_2^0},
\end{equation}
with $SU(2)_L$ transforming vertically and $SU(2)_R$ horizontally. 
Without $U(1)_S$ 
as a gauge symmetry, the model is free of anomalies without the addition 
of the $\psi$ and $\chi$ fermions.
\begin{table}[htb]
\caption{Particle content of proposed model of dark gauge $U(1)$ symmetry.}
\begin{center}
\begin{tabular}{|c|c|c|c|c|c|}
\hline
particles & $SU(3)_C$ & $SU(2)_L$ & $SU(2)_R$ & $U(1)_X$ & $U(1)_S$ \\
\hline
$(u,d)_L$ & 3 & 2 & 1 & 1/6 & 0 \\
$(u,h)_R$ & $3$ & 1 & 2 & 1/6 & $-1/2$ \\
$d_R$ & $3$ & 1 & 1 & $-1/3$ & 0 \\
$h_L$ & 3 & 1 & 1 & $-1/3$ & $-1$ \\ 
$(\nu,l)_L$ & 1 & 2 & 1 & $-1/2$ & 0 \\
$(n,l)_R$ & 1 & 1 & 2 & $-1/2$ & 1/2 \\ 
$\nu_R$ & 1 & 1 & 1 & $0$ & 0 \\
$n_L$ & 1 & 1 & 1 & 0 & 1 \\
\hline
$(\phi_L^+,\phi_L^0)$ & 1 & 2 & 1 & $1/2$ & $0$ \\
$(\phi_R^+,\phi_R^0)$ & 1 & 1 & 2 & 1/2 & 1/2 \\ 
$\eta$ & 1 & 2 & 2 & $0$ & $-1/2$ \\
$\zeta$ & 1 & 1 & 1 & $0$ & $1$ \\
\hline
$(\psi_1^0,\psi_1^-)_R$ & 1 & 1 & 2 & $-1/2$ & 2 \\  
$(\psi_2^+,\psi_2^0)_R$ & 1 & 1 & 2 & $1/2$ & 1 \\  
$\chi_R^+$ & 1 & 1 & 1 & 1 & $-3/2$ \\ 
$\chi_R^-$ & 1 & 1 & 1 & $-1$ & $-3/2$ \\ 
$\chi^0_{1R}$ & 1 & 1 & 1 & 0 & $-1/2$ \\
$\chi^0_{2R}$ & 1 & 1 & 1 & 0 & $-5/2$ \\
\hline
$\sigma$ & 1 & 1 & 1 & 0 & 3 \\
\hline
\end{tabular}
\end{center}
\end{table}
In the presence of gauge $U(1)_S$, the additional anomaly-free conditions 
are all satisfied by the addition of the $\psi$ and $\chi$ fermions. 
The $[SU(3)_C]^2 U(1)_S$ anomaly is canceled between $(u,h)_R$ and $h_L$; 
the $[SU(2)_L]^2 U(1)_S$ anomaly is zero because $(u,d)_L$ and $(\nu,l)_L$ 
do not transform under $U(1)_S$; the $[SU(2)_R]^2 U(1)_S$ and 
$[SU(2)_R]^2 U(1)_X$ anomalies are both canceled by summing over $(u,h)_R$, 
$(n,l)_R$, $(\psi_1^0,\psi_1^-)_R$, and $(\psi_2^+,\psi_2^0)_R$;
the addition of $\chi_R^\pm$ renders the $[U(1)_X]^2 U(1)_S$,  
$U(1)_X [U(1)_S]^2$, $[U(1)_X]^3$, and $U(1)_X$ anomalies zero; and the 
further addition of $\chi^0_{1R}$ and $\chi^0_{2R}$ kills both the 
$[U(1)_S]^3$ and $U(1)_S$ anomalies, i.e.
\begin{eqnarray}
0 &=& 3[6(-1/2)^3-3(-1)^3+2(1/2)^3-(1)^3] \nonumber \\ 
&+& 2(2)^3 +2(1)^3 + 2(-3/2)^3 +(-1/2)^3 + (-5/2)^3, \\ 
0 &=& 3[6(-1/2)-3(-1)+2(1/2)-(1)] \nonumber \\ 
&+& 2(2) +2(1) + 2(-3/2) +(-1/2) + (-5/2). 
\end{eqnarray}

Under $T_{3R} + S$, the neutral scalars $\phi_R^0$ and $\eta_2^0$ are zero, 
so that their vacuum expectation values do not break $T_{3R} + S$ which 
remains as a global symmetry.  However, $\langle \sigma \rangle \neq 0$ 
does break $T_{3R} + S$ and gives masses to $\psi^0_{1R} \psi^0_{2R} - 
\psi^-_{1R} \psi^+_{2R}$, $\chi^+_R \chi^-_R$, and $\chi^0_{1R} \chi^0_{2R}$. 
These exotic fermions all have half-integral charges~\cite{km16} under 
$T_{3R} + S$ and 
only communicate with the others with integral charges through $W_R^\pm$, 
$\sqrt{2} Re(\phi^0_R)$, $\zeta$, and the two extra neutral gauge bosons 
beyond the $Z$.  Some explicit Yukawa terms are
\begin{eqnarray}
&& (\psi_{1R}^0 \phi_R^- + \psi_{1R}^- \bar{\phi}_R^0) \chi_R^+, ~~~ 
(\psi_{2R}^+ \phi_R^0 - \psi_{2R}^0 \phi_R^+) \chi_R^-, \\ 
&& (\psi_{1R}^0 \phi_R^0 - \psi_{1R}^- \phi_R^+) \chi^0_{2R}, ~~~ 
(\psi_{2R}^+ \phi_R^- + \psi_{2R}^0 \bar{\phi}_R^0) \chi^0_{1R}. 
\end{eqnarray}
This dichotomy of particle content results in an additional unbroken 
symmetry of the Lagrangian, i.e. discrete $Z_2$ under which the exotic 
fermions are odd.  Hence dark matter has two layers: those with nonzero 
$T_{3R} + S$ and even $Z_2$, i.e. $n, h, W_R^\pm, \phi^\pm_R, \eta_1^\pm, 
\eta_1^0, \bar{\eta}_1^0$, $\zeta$, and the underlying exotic fermions with 
odd $Z_2$.  Without $\zeta$, a global $S'$ symmetry remains.  With $\zeta$, 
because of the $\zeta^3 \sigma^*$ and $\chi^0_{1R} \chi^0_{1R} \zeta$ terms, 
the $S'$ symmetry breaks to $Z_3$.
\begin{table}[htb]
\caption{Particle content of proposed model under $(T_{3R} + S) \times Z_2$.}
\begin{center}
\begin{tabular}{|c|c|c|c|c|}
\hline
particles & gauge $T_{3R} + S$ & global $S'$ & $Z_3$ & $Z_2$ \\
\hline
$u,d,\nu,l$ & 0 & 0 & 1 & + \\
$(\phi_L^+,\phi_L^0), (\eta_2^+,\eta_2^0), \phi_R^0$ & 0 & 0 & 1 & + \\
$n, \phi_R^+, \zeta$ & 1 & 1 & $\omega$ & + \\
$h, (\eta_1^0, \eta_1^-)$ & $-1$ & $-1$ & $\omega^2$ & + \\ 
\hline
$\psi^+_{2R}, \chi^+_R$ & $3/2,-3/2$ & 0 & $1$ & $-$ \\ 
$\psi^-_{1R}, \chi^-_R$ & $3/2,-3/2$ & $0$ & $1$ & $-$ \\
$\psi^0_{1R}, \psi^0_{2R}$ & $5/2,1/2$ & $1,-1$ & $\omega,\omega^2$ & $-$ \\ 
$\chi^0_{1R},\chi^0_{2R}$ & $-1/2,-5/2$ & $1,-1$ & $\omega,\omega^2$ & $-$ \\
\hline
$\sigma$ & 3 & 0 & 1 & + \\
\hline
\end{tabular}
\end{center}
\end{table}

Let
\begin{equation}
\langle \phi^0_L \rangle = v_1, ~~~ \langle \eta^0_2 \rangle = v_2, ~~~ 
\langle \phi^0_R \rangle = v_R, ~~~ \langle \sigma \rangle = v_S, 
\end{equation}
then the $SU(3)_C \times SU(2)_L \times SU(2)_R \times U(1)_X \times U(1)_S$ 
gauge symmetry is broken to $SU(3)_C \times U(1)_Q$ with $S'$, which becomes 
$Z_3$, as shown in Table 2 with $\omega^3 = 1$.  The discrete $Z_2$ 
symmetry is unbroken.  Note that the global $S'$ assignments for 
the exotic fermions are not $T_{3R} + S$ because of $v_S$ which breaks the 
gauge $U(1)_S$ by 3 units.  

\noindent \underline{\it Gauge sector}~:\\
Consider now the masses of the gauge bosons.  The charged ones, $W_L^\pm$ 
and $W_R^\pm$, do not mix because of $S'(Z_3)$, as in the original alternative 
left-right models.  Their masses are given by
\begin{equation}
M_{W_L}^2 = {1 \over 2} g_L^2 (v_1^2 + v_2^2), ~~~ 
M_{W_R}^2 = {1 \over 2} g_R^2 (v_R^2 + v_2^2).
\end{equation}
Since $Q = I_{3L} + I_{3R} + X$, the photon is given by
\begin{equation}
A = {e \over g_L} W_{3L} + {e \over g_R} W_{3R} + {e \over g_X} X,
\end{equation}
where $e^{-2} = g_L^{-2} + g_R^{-2} + g_X^{-2}$.  Let
\begin{eqnarray}
Z &=& (g_L^2 + g_Y^2)^{-1/2} \left( g_L W_{3L} - {g_Y^2 \over g_R} W_{3R} 
- {g_Y^2 \over g_X} X \right), \\ 
Z' &=& (g_R^2 + g_X^2)^{-1/2} ( g_R W_{3R} - g_X X),
\end{eqnarray}
where $g_Y^{-2} = g_R^{-2} + g_X^{-2}$, 
then the $3 \times 3$ mass-squared matrix spanning $(Z,Z',S)$ has the 
entries:
\begin{eqnarray}
M^2_{ZZ} &=& {1 \over 2} (g_L^2 + g_Y^2) (v_1^2 + v_2^2), \\ 
M^2_{Z'Z'} &=& {1 \over 2} (g_R^2 + g_X^2) v_R^2 + {g_X^4 v_1^2 + g_R^4 v_2^2 
\over 2(g_R^2 + g_X^2)}, \\ 
M^2_{SS} &=& 18 g_S^2 v_S^2 + {1 \over 2} g_S^2 (v_R^2 + v_2^2), \\ 
M^2_{ZZ'} &=& {\sqrt{g_L^2 + g_Y^2} \over 2\sqrt{g_R^2 + g_X^2}} (g_X^2 v_1^2 - 
g_R^2 v_2^2), \\ 
M^2_{ZS} &=& {1 \over 2} g_S \sqrt{g_L^2 + g_Y^2} v_2^2, \\ 
M^2_{Z'S} &=& -{1 \over 2} g_S \sqrt{g_R^2 - g_X^2} v_R^2 - 
{g_S g_R v_2^2 \over 2 \sqrt{g_R^2 + g_X^2}}.
\end{eqnarray}
Their neutral-current interactions are given by
\begin{eqnarray}
{\cal L}_{NC} &=& e A_\mu j^\mu_Q + g_Z Z_\mu (j^\mu_{3L} - \sin^2 \theta_W 
j^\mu_Q) \nonumber \\ 
&+& (g_R^2 + g_X^2)^{-1/2} Z'_\mu (g_R^2 j^\mu_{3R} - g_X^2 j^\mu_X) + 
g_S S_\mu j^\mu_S,
\end{eqnarray}
where $g_Z^2 = g_L^2 + g_Y^2$ and $\sin^2 \theta_W = g_Y^2/g_Z^2$.

In the limit $v^2_{1,2} << v^2_R, v^2_S$, the mass-squared matrix spanning 
$(Z',S)$ may be simplified if we assume
\begin{equation}
{v_S^2 \over v_R^2} = {(g_R^2 + g_X^2 + g_S^2)^2 \over 36 g_S^2 
(g_R^2 + g_X^2 - g_S^2)},
\end{equation}
and let
\begin{equation}
\tan \theta_D = {\sqrt{g_R^2 + g_X^2} - g_S \over \sqrt{g_R^2 + g_X^2} + g_S}, 
\end{equation}
then
\begin{equation}
\pmatrix{D_1 \cr D_2} = \pmatrix{\cos \theta_D & \sin \theta_D \cr 
-\sin \theta_D & \cos \theta_D} \pmatrix{Z' \cr S},
\end{equation}
with mass eigenvalues given by
\begin{eqnarray}
M^2_{D_1} &=&  \sqrt{g_R^2 + g_X^2} \sqrt{g_R^2 + g_X^2 + g_S^2} 
{v_R^2 \over 2 \sqrt{2} \cos \theta_D}, \\ 
M^2_{D_2} &=&  \sqrt{g_R^2 + g_X^2} \sqrt{g_R^2 + g_X^2 + g_S^2} 
{v_R^2 \over 2 \sqrt{2} \sin \theta_D}.
\end{eqnarray}
In addition to the assumption of Eq.~(18), let us take for example 
\begin{equation}
2 g_S = \sqrt{g_R^2 + g_X^2},
\end{equation}
then $\sin \theta_D = 1/\sqrt{10}$ and $\cos \theta_D = 3/\sqrt{10}$. 
Assuming also that $g_R = g_L$, we obtain
\begin{eqnarray}
&& {g_X^2 \over g_Z^2} = {\sin^2 \theta_W \cos^2 \theta_W \over \cos 2 \theta_W}, 
~~~ {g_S \over g_Z} = {\cos^2 \theta_W \over 2 \sqrt{\cos 2 \theta_W}}, \\ 
&& {v^2_S \over v^2_R} = {25 \over 108}, ~~~ M^2_{D_2} = 3 M^2_{D_1} = 
{5 \cos^4 \theta_W \over 4 \cos 2 \theta_W} g_Z^2 v_R^2.
\end{eqnarray}
The resulting gauge interactions of $D_{1,2}$ are given by
\begin{eqnarray}
{\cal L}_D &=& {g_Z \over \sqrt{10} \sqrt{\cos 2 \theta_W}} \{ [3 \cos 2 
\theta_W j^\mu_{3R} - 3 \sin^2 \theta_W j^\mu_X + (1/2)\cos^2 \theta_W j^\mu_S] 
D_{1\mu} \nonumber \\ && ~~~~~~~~~~~~~ + [- \cos 2 \theta_W j^\mu_{3R} + 
\sin^2 \theta_W j^\mu_X + (3/2)\cos^2 \theta_W j^\mu_S] D_{2\mu} \}.
\end{eqnarray}
Since $D_2$ is $\sqrt{3}$ times heavier than $D_1$ in this example, the latter 
would be produced first in $pp$ collisions at the Large Hadron Collider 
(LHC).

\noindent \underline{\it Fermion sector}~:\\
All fermions obtain masses through the four vacuum expectation values 
of Eq.~(6) except $\nu_R$ which is allowed to have an invariant Majorana 
mass.  This means that neutrino masses may be small from the usual 
canonical seesaw mechanism.  The various Yukawa terms for the quark and 
lepton masses are
\begin{eqnarray}
-{\cal L}_Y &=& {m_u \over v_2} [\bar{u}_R ({u}_L {\eta}^0_2 - {d}_L 
\eta_2^+) + \bar{h}_R(-{u}_L \eta_2^- + {d}_L {\eta}_1^0)] 
\nonumber \\ 
&+& {m_d \over v_1} (\bar{u}_L \phi_L^+ + \bar{d}_L \phi_L^0)d_R + 
{m_h \over v_R} (\bar{u}_R \phi_R^+ + \bar{h}_R \phi_R^0)h_L \nonumber \\ 
&+& {m_l \over v_2} [(\bar{\nu}_L \eta_1^0 + \bar{l}_L \eta_1^-)n_R + 
(\bar{\nu}_L \eta_2^+ + \bar{l}_L \eta_2^0) l_R] \nonumber \\ 
&+& {m_D \over v_1} \bar{\nu}_R({\nu}_L {\phi}_L^0 - {l}_L \phi_L^+) + 
{m_n \over v_R} \bar{n}_L (n_R \phi_R^0 - l_R \phi_R^-) + H.c.
\end{eqnarray}
These terms show explicitly that the assignments of Tables 1 and 2 are 
satisfied.

As for the exotic $\psi$ and $\chi$ fermions, they have masses from the 
Yukawa terms of Eqs.~(4) and (5), as well as
\begin{equation}
(\phi^0_{1R} \psi^0_{2R} - \psi^-_{1R} \psi^+_{2R}) \sigma^*, ~~~ 
\chi^-_R \chi^+_R \sigma, ~~~ \chi^0_{1R} \chi^0_{2R} \sigma.
\end{equation}
As a result, two neutral Dirac fermions are formed from the matrix 
linking $\chi^0_{1R}$ and $\psi^0_{1R}$ to $\chi^0_{2R}$ and $\psi^0_{2R}$. 
Let us call the lighter of these two Dirac fermions $\chi_0$, then it 
is one component of dark matter of our model.  The other will be the 
scalar $\zeta$, to be discussed later.  Note that $\chi_0$ communicates 
with $\zeta$ through the allowed $\chi^0_{1R} \chi^0_{1R} \zeta$ 
interaction.  Note also that the allowed Yukawa terms
\begin{equation}
\bar{d}_R h_L \zeta, ~~~ \bar{n}_L \nu_R \zeta
\end{equation}
enable the dark fermions $h$ and $n$ to decay into $\zeta$.

\noindent \underline{\it Scalar sector}~:\\
Consider the most general scalar potential consisting of $\Phi_{L,R}$, $\eta$, 
and $\sigma$. Let
\begin{equation}
\eta = \pmatrix{\eta_1^0 & \eta_2^+ \cr \eta_1^- & \eta_2^0}, ~~~ 
\tilde{\eta} = \sigma_2 \eta^* \sigma_2 = \pmatrix{\bar{\eta}_2^0 & 
-\eta_1^+ \cr -\eta_2^- & \bar{\eta}_1^0},
\end{equation}
then
\begin{eqnarray}
V &=& -\mu^2_L \Phi_L^\dagger \Phi_L - \mu^2_R \Phi_R^\dagger \Phi_R - 
\mu^2_\sigma \sigma^* \sigma - \mu^2_\eta Tr(\eta^\dagger \eta) 
+ [\mu_3 \Phi_L^\dagger \eta \Phi_R + H.c.] \nonumber \\ 
&+& {1 \over 2} \lambda_L (\Phi_L^\dagger \Phi_L)^2 + {1 \over 2} \lambda_R 
(\Phi_R^\dagger \Phi_R)^2 + {1 \over 2} \lambda_\sigma (\sigma^* \sigma)^2 + 
{1 \over 2} \lambda_\eta [Tr(\eta^\dagger \eta)]^2 + {1 \over 2} \lambda'_\eta 
Tr(\eta^\dagger \eta \eta^\dagger \eta) \nonumber \\ 
&+& \lambda_{LR} (\Phi_L^\dagger \Phi_L)(\Phi_R^\dagger \Phi_R) + \lambda_{L\sigma} 
(\Phi_L^\dagger \Phi_L) (\sigma^* \sigma) + \lambda_{R\sigma} (\Phi_R^\dagger 
\Phi_R) (\sigma^* \sigma) + \lambda_{\sigma \eta} (\sigma^* \sigma) Tr(\eta^\dagger 
\eta) \nonumber \\ 
&+& \lambda_{L\eta} \Phi_L^\dagger \eta \eta^\dagger \Phi_L + \lambda'_{L\eta} 
\Phi_L^\dagger \tilde{\eta} \tilde{\eta}^\dagger \Phi_L + \lambda_{R\eta} 
\Phi_R^\dagger \eta^\dagger \eta \Phi_R + \lambda'_{R\eta} 
\Phi_R^\dagger \tilde{\eta}^\dagger \tilde{\eta} \Phi_R.
\end{eqnarray}
Note that
\begin{eqnarray}
2 |det(\eta)|^2 &=& [Tr(\eta^\dagger \eta)]^2 - Tr(\eta^\dagger \eta \eta^\dagger 
\eta), \\ 
(\Phi_L^\dagger \Phi_L) Tr(\eta^\dagger \eta) &=& \Phi_L^\dagger \eta \eta^\dagger 
\Phi_L + \Phi_L^\dagger \tilde{\eta} \tilde{\eta}^\dagger \Phi_L, \\ 
(\Phi_R^\dagger \Phi_R) Tr(\eta^\dagger \eta) &=& \Phi_R^\dagger \eta^\dagger \eta 
\Phi_R + \Phi_R^\dagger \tilde{\eta}^\dagger \tilde{\eta} \Phi_L.
\end{eqnarray} 
The minimum of $V$ satisfies the conditions
\begin{eqnarray}
\mu_L^2 &=& \lambda_L v_1^2 + \lambda_{L\eta} v_2^2 + \lambda_{LR} v_R^2 + 
\lambda_{L\sigma} v_S^2 + \mu_3 v_2 v_R/v_1, \\ 
\mu_\eta^2 &=& (\lambda_\eta + \lambda'_\eta) v_2^2 + \lambda_{L\eta} v_1^2 + 
\lambda_{R\eta} v_R^2 + \lambda_{\sigma\eta} v_S^2 + \mu_3 v_1 v_R/v_2, \\ 
\mu_R^2 &=& \lambda_R v_R^2 + \lambda_{LR} v_1^2 + \lambda_{R\eta} v_2^2 + 
\lambda_{R\sigma} v_S^2 + \mu_3 v_1 v_2/v_R, \\ 
\mu_\sigma^2 &=& \lambda_\sigma v_S^2 + \lambda_{L\sigma} v_1^2 + 
\lambda_{\sigma\eta} v_2^2 + \lambda_{R\sigma} v_R^2. 
\end{eqnarray}
The $4 \times 4$ mass-squared matrix spanning $\sqrt{2}Im(\phi_L^0,\eta_2^0,
\phi_R^0,\sigma)$ is then given by
\begin{equation}
{\cal M}^2_I = \mu_3 \pmatrix{-v_2v_R/v_1 & v_R & v_2 & 0 \cr v_R & -v_1v_R/v_2 
& v_1 & 0 \cr v_2 & v_1 & -v_1v_2/v_R & 0 \cr 0 & 0 & 0 & 0}.
\end{equation}
and that spanning $\sqrt{2}Re(\phi_L^0,\eta_2^0,\phi_R^0,\sigma)$ is 
\begin{equation}
{\cal M}^2_R = {\cal M}^2_I + 2\pmatrix{ \lambda_L v_1^2 & \lambda_{L\eta}v_1v_2 &  
\lambda_{LR}v_1 v_R &  \lambda_{L\sigma} v_1 v_S \cr \lambda_{L\eta}v_1v_2 
& (\lambda_\eta + \lambda'_\eta)v_2^2 & \lambda_{R\eta}v_2v_R & 
\lambda_{\sigma\eta} v_2 v_S \cr  \lambda_{LR} v_1 v_R &  \lambda_{R\eta} 
v_2 v_R &  \lambda_R v_R^2 &  \lambda_{R\sigma} v_R v_S \cr 
 \lambda_{L\sigma} v_1v_S &  \lambda_{\sigma\eta}v_2v_S & 
\lambda_{R\sigma} v_R v_S &  \lambda_\sigma v_S^2}. 
\end{equation}
Hence there are three zero eigenvalues in ${\cal M}^2_I$ with one nonzero 
eigenvalue $-\mu_3[v_1 v_2/v_R + v_R (v_1^2+v_2^2)/v_1v_2]$ corresponding 
to the eigenstate $(-v_1^{-1},v_2^{-1},v_R^{-1},0)/\sqrt{v_1^{-2}+v_2^{-2}+v_R^{-2}}$. 
In ${\cal M}^2_R$, the linear combination $H = (v_1,v_2,0,0)/\sqrt{v_1^2+v_2^2}$, 
is the standard-model Higgs boson, with 
\begin{equation}
m_H^2 = 2[\lambda_L v_1^4 + (\lambda_\eta + \lambda'_\eta) v_2^4 + 2 \lambda_{L\eta} 
v_1^2 v_2^2]/(v_1^2 + v_2^2).
\end{equation}
The other three scalar bosons are much heavier, with suppressed mixing to $H$, 
which may all be assumed to be small enough to avoid the constraints from 
dark-matter direct-search experiments.
The addition of the scalar $\zeta$ introduces two important new terms:
\begin{equation}
\zeta^3 \sigma^*, ~~~ (\eta_1^0 \eta_2^0 - \eta_1^- \eta_2^+) \zeta.
\end{equation}
The first term breaks global $S'$ to $Z_3$, and the second term mixes $\zeta$ 
with $\eta_1^0$ through $v_2$.  We assume the latter to be negligible, so 
that the physical dark scalar is mostly $\zeta$.

\noindent \underline{\it Present phenomenological constraints}~:\\
Many of the new particles of this model interact with those of the 
standard model.  The most important ones are the neutral $D_{1,2}$ gauge 
bosons, which may be produced at the LHC through their couplings to $u$ 
and $d$ quarks, and decay to charged leptons ($e^-e^+$ and $\mu^-\mu^+$). 
As noted previously, in our chosen example, $D_1$ is the lighter of the 
two.  Hence current search limits for a $Z'$ boson are 
applicable~\cite{atlas14,cms14}. 
The $c_{u,d}$ coefficients used in the data analysis are
\begin{equation}
c_u = (g_{uL}^2 + g_{uR}^2) B = 0.0273~B, ~~~ 
c_d = (g_{dL}^2 + g_{dR}^2) B = 0.0068~B,
\end{equation}
where $B$ is the branching fraction of $Z'$ to $e^-e^+$ and $\mu^-\mu^+$.
Assuming that $D_1$ decays to all the particles listed in Table 2, except 
for the scalars which become the longitudinal components of the various 
gauge bosons, we find $B = 1.2 \times 10^{-2}$.  Based on the 2016 LHC 13 TeV 
data set, this translates to a bound of about 4 TeV on the $D_1$ mass.

The would-be dark-matter candidate $n$ is a Dirac fermion which couples 
to $D_{1,2}$ which also couples to quarks.  Hence severe limits exist on 
the masses of $D_{1,2}$ from underground direct-search experiments as well. 
The annihilation cross section of $n$ through $D_{1,2}$ would then be 
too small, so that its relic abundance would be too big for it to be 
a dark-matter candidate.  Its annihilation at rest through 
$s$-channel scalar exchange is $p$-wave suppressed and does not help. 
As for the $t$-channel diagrams, they also turn out to be too small. 
Previous studies where $n$ is chosen as dark matter are now ruled out.

\noindent \underline{\it Dark sector}~:\\
Dark matter is envisioned to have two components.  One is a Dirac fermion 
$\chi_0$ which is a mixture of the four neutral fermions of odd $Z_2$, 
and the other is a complex scalar boson which is mostly $\zeta$.  
The annihilation $\chi_0 \bar{\chi}_0 \to \zeta \zeta^*$ determines 
the relic abundance of $\chi_0$, and the annihilation $\zeta \zeta^* \to 
H H$, where $H$ is the standard-model Higgs boson, determines that of 
$\zeta$.  The direct $\zeta \zeta^* H$ coupling is assumed small to 
avoid the severe constraint in direct-search experiments.

Let the interaction of $\zeta$ with $\chi_0$ be $f_0 \zeta \chi_{0R} \chi_{0R} 
+ H.c.$, then the annihilation cross section of $\chi_0 \bar{\chi}_0$ to 
$\zeta \zeta^*$ times relative velicity is given by
\begin{equation}
\langle \sigma \times v_{rel} \rangle_\chi = {f_0^4 \over 4 \pi m_{\chi_0}} 
{(m^2_{\chi_0} - m^2_\zeta)^{3/2} \over (2 m^2_{\chi_o} - m^2_\zeta)^2}.
\end{equation}

Let the effective interaction strength of $\zeta \zeta^*$ with $H H$ be 
$\lambda_0$, then the annihilation cross section of $\zeta \zeta^*$ 
to $H H$ times relative velicity is given by
\begin{equation}
\langle \sigma_\zeta \times v_{rel} \rangle_\zeta = {\lambda_0^2 \over 16 \pi} 
{(m_\zeta^2 - m_H^2)^{1/2} \over m^3_\zeta}.
\end{equation}
Note that $\lambda_0$ is the sum over several interactions.  The quartic 
coupling $\lambda_{\zeta H}$ is assumed negligible, to suppress the trilinear 
$\zeta \zeta^* H$ coupling which contributes to the elastic $\zeta$ scattering 
cross section off nuclei.  However, the trilinear couplings 
$\zeta \zeta^* Re(\phi_R^0)$ and $Re(\phi_R^0) H H$  are proportional to $v_R$, 
and the trilinear couplings $\zeta \zeta^* Re(\sigma)$ and $Re(\sigma) H H$  
are proportional to $v_S$.  Hence their effective contributions to $\lambda_0$ 
are proportional to $v_R^2/m^2[\sqrt{2}Re(\phi_R^0)]$ and $v_S^2/m^2[\sqrt{2}
Re(\sigma)]$, which are not suppressed.

\begin{figure}[htb]
\centering
\vspace*{0.5cm}
\includegraphics[scale=1.25]{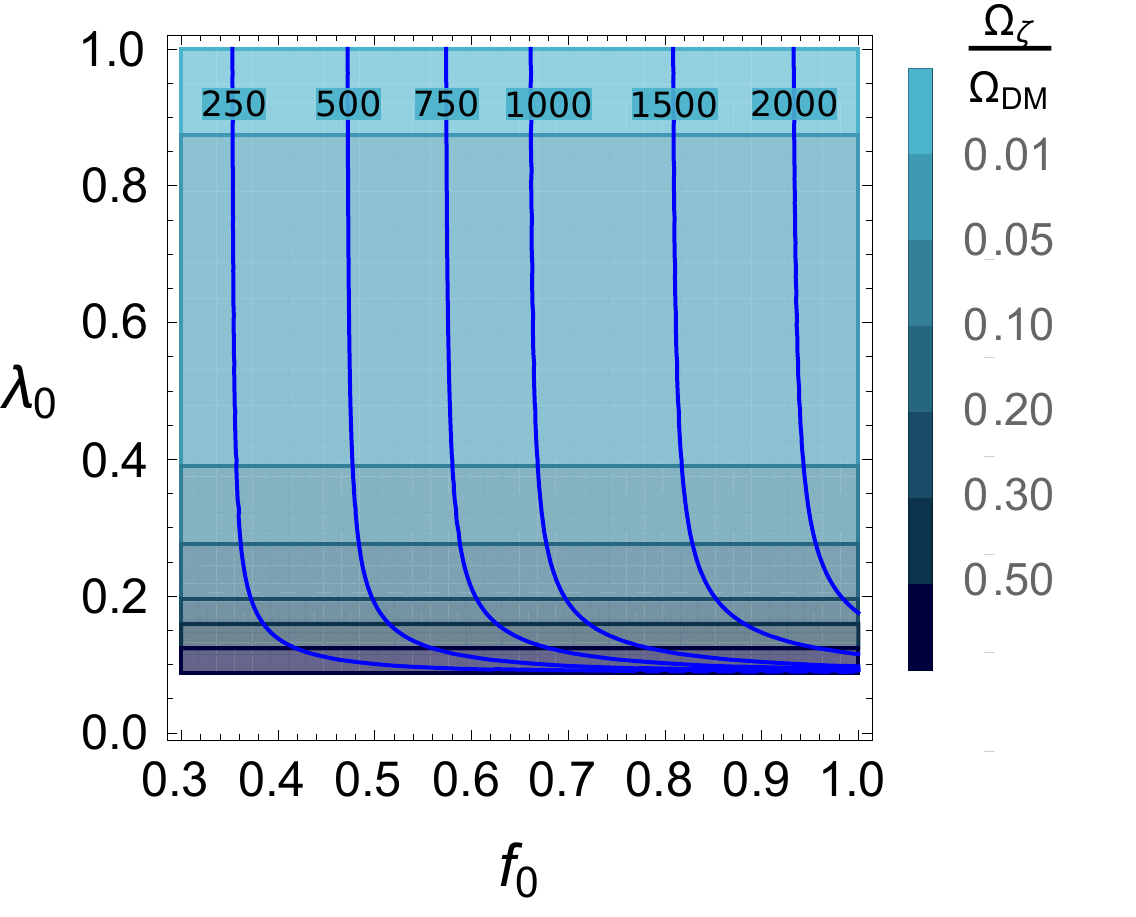}
\caption{Relic-abundance constraints on $\lambda_0$ and $f_0$ for $m_\zeta = 
150$ GeV and various values of $m_{\chi_0}$. }
\end{figure}
As a rough estimate, we will assume that 
\begin{equation}
\langle \sigma \times v_{rel} \rangle_\chi^{-1} + 
\langle \sigma_\zeta \times v_{rel} \rangle_\zeta^{-1}
= (4.4 \times 10^{-26}~cm^3/s)^{-1}
\end{equation}
to satisfy the condition of dark-matter relic abundance~\cite{sdb12} of 
the Universe. 
For given values of $m_\zeta$ and $m_{\chi_0}$, the parameters 
$\lambda_0$ and $f_0$ are thus constrained.  We show in Fig.~1 the plots of 
$\lambda_0$ versus $f_0$ for $m_\zeta = 150$ GeV and various values of 
$m_{\chi_0}$.
Since $m_\zeta$ is fixed at 150 GeV, $\lambda_0$ is also fixed for a given 
fraction of $\Omega_\zeta/\Omega_{DM}$.  To adjust for the rest of dark matter, 
$f_0$ must then vary as a function of $m_{\chi_0}$ according to Eq.~(44). 

As for direct detection, both $\chi_0$ and $\zeta$ have possible interactions 
with quarks through the gauge bosons $D_{1,2}$ and the standard-model 
Higgs boson $H$.  They are suppressed by making the $D_{1,2}$ masses heavy, 
and the $H$ couplings to $\chi_0$ and $\zeta$ small.  In our example with 
$m_\zeta = 150$ GeV, let us choose $m_{\chi_0} = 500$ Gev and the relic 
abundances of both to be equal.  From Fig.~1, these choices translate to 
$\lambda_0 = 0.12$ and $f_0 = 0.56$.

Consider first the $D_{1,2}$ interactions. 
Using Eq.~(26), we obtain
\begin{eqnarray}
&& g^V_u(D_1) = 0.0621, ~~~ g^V_d(D_1) = 0.0184, ~~~ g_\zeta(D_1) = 0.1234, \\
&& g^V_u(D_2) = -0.1235, ~~~ g^V_d(D_2) = -0.0062, ~~~ g_\zeta(D_2) = 0.3701.
\end{eqnarray}
The effective $\zeta$ elastic scattering cross section through $D_{1,2}$ is 
then completely determined as a function of the $D_1$ mass (because 
$M_{D_2} = \sqrt{3} M_{D_1}$ in our example), i.e.
\begin{equation}
{\cal L}^V_{\zeta q} = {(\zeta^* \partial_\mu - \zeta \partial_\mu \zeta^*) \over 
M^2_{D_1}} [(-7.57 \times 10^{-3}) \bar{u} \gamma^\mu u + (1.51 \times 10^{-3}) 
\bar{d} \gamma^\mu d].
\end{equation}
Using the latest LUX result~\cite{lux17} and Eq.~(25), we obtain 
$v_R > 35$ TeV which translates to $M_{D_1} > 18$ TeV, and $M_{W_R} > 16$ 
TeV. 

The $\bar{\chi_0} \gamma_\mu \chi_0$ couplings to $D_{1,2}$ depend on the 
$2 \times 2$ mass matrix linking $(\chi_1,\psi_1)$ to $(\chi_2,\psi_2)$ 
which has two mixing angles and two mass eigenvalues, the lighter one 
being $m_{\chi_0}$.  By adjusting these parameters, it is possible to make 
the effective $\chi_0$ interaction with xenon negligibly small.  Hence 
there is no useful limit on the $D_1$ mass in this case.

Direct search also constrains the coupling of the Higgs boson to $\zeta$ 
(through a possible trilinear $\lambda_{\zeta H} \sqrt{2} v_H \zeta^* \zeta$ 
interaction) or $\chi_0$ (through an effective Yukawa coupling $\epsilon$ 
from $H$ mixing with $\sigma_R$ and $\phi_R^0$).  Let their effective 
interactions with quarks through $H$ exchange be given by
\begin{equation}
{\cal L}^S_{\zeta q} = {\lambda_{\zeta H} m_q \over m_H^2} \zeta^* \zeta 
\bar{q} q + {\epsilon f_q \over m_H^2} \bar{\chi}_0 \chi_0 \bar{q} q, 
\end{equation}
where $f_q = m_q/\sqrt{2}v_H = m_q/(246~{\rm GeV})$.  The spin-independent 
direct-detection cross section per nucleon in the former is given by
\begin{equation}
\sigma^{SI} = {\mu^2_\zeta \over \pi A^2} [\lambda_p Z + (A-Z)\lambda_n]^2,
\end{equation}
where $\mu_{\zeta} = m_\zeta M_A/(m_\zeta + M_A)$ is the reduced mass of 
the dark matter, and~\cite{bbps09}
\begin{equation}
\lambda_N = \left[ \sum_{u,d,s} f_q^N + {2 \over 27} \left( 1 - \sum_{u,d,s} 
f_q^N \right) \right] {\lambda_{\zeta H} m_N \over 2 m_\zeta m_H^2},
\end{equation}
with~\cite{jlqcd08}
\begin{eqnarray}
&& f_u^p = 0.023, ~~~ f_d^p = 0.032, ~~~ f_s^p = 0.020, \\ 
&& f_u^n = 0.017, ~~~ f_d^n = 0.041, ~~~ f_s^n = 0.020.
\end{eqnarray}
For $m_\zeta = 150$ GeV, we have
\begin{equation}
\lambda_p = 2.87 \times 10^{-8} \lambda_{\zeta H}~{\rm GeV}^{-2}, ~~~ 
\lambda_n = 2.93 \times 10^{-8} \lambda_{\zeta H}~{\rm GeV}^{-2}. 
\end{equation}
Using $A=131$, $Z=54$, and $M_A=130.9$ atomic mass units for the LUX 
experiment~\cite{lux17}, and twice the most recent bound of 
$2 \times 10^{-46}~cm^2$ (because $\zeta$ is assumed to account for 
only half of the dark matter) at this mass, we find
\begin{equation}
\lambda_{\zeta H} < 9.1 \times 10^{-4}.
\end{equation}
As noted earlier, this is negligible for considering the annihilation 
cross section of $\zeta$ to $H$.

For the $H$ contribution to the $\chi_0$ elastic cross section off nuclei, 
we replace $m_\zeta$ with $m_{\chi_0} = 500$ GeV in Eq.~(51) and 
$\lambda_{\zeta H}/2 m_\zeta$ with $\epsilon/\sqrt{2}v_H$ in Eq.~(52). 
Using the experimental data at 500 GeV, we obtain the bound.
\begin{equation}
\epsilon < 9.6 \times 10^{-4}.
\end{equation}
From the above discussion, it is clear that our model allows for the 
discovery of dark matter in direct-search experiments in the future 
if these bounds are only a little above the actual values of 
$\lambda_{\zeta H}$ and $\epsilon$.

\noindent \underline{\it Conclusion and outlook}~:~
In the context of the alternative left-right model, a new gauge $U(1)_S$ 
symmetry has been proposed to stabilize dark matter.  This is accomplished 
by the addition of a few new fermions to cancel all the gauge anomalies, 
as shown in Table 1.  As a result of this particle content, an automatic 
unbroken $Z_2$ symmetry exists on top of $U(1)_S$ which is broken to a 
conserved residual $Z_3$ symmetry.  Thus dark matter has two components. 
One is the Dirac fermion $\chi_0 \sim (\omega,-)$ and the other the 
complex scalar $\zeta \sim (\omega,+)$ under $Z_3 \times Z_2$.  We have 
shown how they may account for the relic abundance of dark matter in the 
Universe, and satisfy present experimental search bounds.

Whereas we have no specific prediction for discovery in direct-search 
experiments, our model will be able to accommodate any positive result 
in the future, just like many other existing proposals.  To single out 
our model, many additional details must also be confirmed.  Foremost 
are the new gauge bosons $D_{1,2}$.  Whereas the LHC bound is about 4 TeV, 
the direct-search bound is much higher provided that $\zeta$ is a 
significant fraction of dark matter.  If $\chi_0$ dominates instead, 
the adjustment of free parameters of our model can lower this bound 
to below 4 TeV.  In that case, future $D_{1,2}$ observations are still 
possible at the LHC as more data become available.

Another is the exotic $h$ quark which is easily produced if kinematically 
allowed.  It would decay to $d$ and $\zeta$ through the direct 
$\bar{d}_R h_L \zeta$ coupling of Eq.~(29).  Assuming that this branching 
fraction is 100\%, the search at the LHC for 2 jets plus missing energy 
puts a limit on $m_h$ of about 1.0 TeV, as reported by the CMS 
Collaboration~\cite{cms17} based on the $\sqrt{s} = 13$ TeV data at 
the LHC with an integrated luminosity of 35.9 fb$^{-1}$ for a single 
scalar quark.  

If the $\bar{d}_R h_L \zeta$ coupling is very small, then $h$ may also 
decay significantly to $u$ and a virtual $W_R^-$, with $W_R^-$ becoming 
$\bar{n} l^-$, and $\bar{n}$ becoming $\bar{\nu} \zeta^*$. 
This has no analog in the usual searches for supersymmetry or the 
fourth family because $W_R$ is heavy ($> 16$ TeV). 
To be specific, the final states of 2 jets plus $l_1^- l_2^+$ 
plus missing energy should be searched for.  As more data 
are accumulated at the LHC, such events may become observable.

\noindent \underline{\it Acknowledgement}~:~
This work was supported in part by the U.~S.~Department of Energy Grant 
No. DE-SC0008541.

\baselineskip 18pt
\bibliographystyle{unsrt}

\end{document}